\documentclass[10pt,superscriptaddress,twocolumn,amsmath,amssymb,aps,pra,showpacs]{revtex4-1}
\usepackage{mathrsfs}
\usepackage{graphicx}
\usepackage{dcolumn}
\usepackage{bm}
\usepackage[title,titletoc,header]{appendix}

\usepackage{amssymb}
\usepackage{amsmath}
\usepackage{mathrsfs}
\usepackage{amsmath}
\usepackage{subfigure}
\usepackage{float}
\usepackage[title,titletoc,header]{appendix}
\usepackage{graphicx}

\newcommand{\be}{\begin{equation}}
\newcommand{\ee}{\end{equation}}
\newcommand{\bea}{\begin{eqnarray}}
\newcommand{\eea}{\end{eqnarray}}

\begin{document}

\title{Energy bands in a three dimension simple cubic lattice of contact potential}
\author{Yi-Cai Zhang}\thanks{ E-mail:~
zhangyicai123456@163.com}
\affiliation{School of Physics and Materials Science, Guangzhou University, Guangzhou 510006, China}

\author{J. M. Zhang}\thanks{Corresponding author. E-mail:~wdlang06@163.com
}
\affiliation{Fujian Provincial Key Laboratory of Quantum Manipulation and New Energy Materials, College of Physics and Energy,
Fujian Normal University, Fuzhou 350007, China}
\affiliation{Fujian Provincial Collaborative Innovation Center for Optoelectronic Semiconductors and Efficient Devices, Xiamen, 361005, China}


\date{\today}
\begin{abstract}
In this work, we investigate energy bands in a three dimensional simple cubic lattice of contact potential.
The energy bands in the first  Brillouin Zone are obtained with Ewald's summation method.
 In comparison with single point potential, the presence of  lattice potential changes the existence condition of negative energy states near zero energy.
 It is found that the system always has negative energy states for an arbitrarily weak periodic potential.
In addition, we prove that if an irreducible unitary representation is not a trivial representation of group of wave vector,
 the corresponding wave functions at lattice sites would be zero. With this theorem,  the degeneracy of energy bands is explained with group theory.
%
 Furthermore, we find that there exists some energy bands which are not affected by the lattice potential. We call their corresponding eigenstates as dark states.
 The physical mechanism of the dark states is explained by explicitly constructing the standing wave-type Bloch wave functions.

\end{abstract}

\maketitle
\section{Introduction}
The notion of energy band is a cornerstone of modern solid state theory. It is now an indispensable part of
undergraduate solid state courses. Unfortunately, there
are very few models which are amiable to simple analytic or numerical approaches so that can serve pedagogical purposes. In current textbooks, the most commonly
used model is the one dimensional Kronig-Penney model consisting of an infinite periodic array of rectangular
potential barriers \cite{Lifshitz1980}. The Hamiltonian is
\begin{align}
H_{1D}=-\frac{\hbar^2\partial^2}{2m\partial x^2}+\sum_{n=-\infty}^{\infty}V_{1D}(x-nd),
\end{align}
 where $d$ is the lattice constant, and the atomic potential is
\begin{align}
V_{1D}(x)=h\theta(w/2-|x|),
\end{align}
which is a rectangular potential of height $h$ and width $w$.
Note that here and henceforth $\theta(.)$ denotes the Heaviside
step function. Sometimes, the model is further simplified
or idealized by taking the limit of $w\rightarrow0$, $h\rightarrow\infty$ with
their product $hw=g_{1D} $ fixed, so as to reduce the rectangular potential to a Dirac delta potential. The Hamiltonian
is now
\begin{align}\label{Penny}
H_{1D}=-\frac{\hbar^2\partial^2}{2m\partial x^2}+\sum_{n=-\infty}^{\infty}g_{1D} \delta(x-nd),
\end{align}
where $g_{1D}$ is the one-dimension delta potential strength.

In this paper, we propose to study the energy bands of a three dimensional analog of the Kronig-Penney
model. That is, we take a cubic lattice and put a zero-ranged contact potential at each lattice site. However, it
is crucial that the contact potential cannot be a three dimensional Dirac delta potential, the straightforward generalization of the one dimensional Dirac delta potential.
The reason is as follows. The delta potential may be viewed as a spherical square well (barrier) with an infinitesimal width $w$ and infinitely large strength $h$ which is inversely proportional cube of width, i.e, $h\propto 1/w^3$.  However, it is found that in the limit of $w\rightarrow0$, the repulsive potential barrier ($h>0$) has no effect on a scattering wave function, e.g., the scattering phase shift is always vanishing \cite{Atkinson1975}. For potential well ($h<0$), the phase shift has no well-defined limit as $w\rightarrow0$ due to the infinite number oscillations of the wave function inside potential well.
In order to get a meaningful scattering amplitude, the three dimensional delta potential usually needs
regularization and renormalization \cite{HUANG1989,Jackiw1991}.

On the other hand, a well-defined s-wave scattering amplitude can be also obtained by a Huang-Yang pseudo-potential  \cite{Kerson1957,Huang1987}, namely
 \begin{align}
V(\mathbf{r})=g\delta^3(\mathbf{r})\frac{\partial }{\partial |\mathbf{r}|}(|\mathbf{r}|.),
    	\label{1}
\end{align}
where the potential strength parameter $g=\frac{4\pi\hbar^2a_{3D}}{2m}$,  $a_{3D} $ is the s-wave scattering length in three dimension, $m$ is particle mass, and $\hbar$ is reduced Plank constant.
At low energy limit,
namely, $\sqrt{E}r_0\ll1$  (where $E$ is energy of incident wave, and $r_0$ is the force range of true potential), the s-wave scattering is dominant and the scattering of higher partial waves can be neglected.
In such a case, the Huang-Yang potential can be used to replace the true potential in scattering problem. In fact, the two methods, i.e., the renormalization procedure for delta potential and  the use of the Huang-Yang pseudopotential, are equivalent, in the sense that they can give same physical results, e.g., the same s-wave scattering amplitudes. Furthermore, such a model potential has important applications in the current cold atom physics \cite{Ueda,Pitaevskii2016}.

The effect of the pseudopotential on the wave function
can be seen from the eigenvalue equation [$E = \hbar^2q^2/(2m$)]
\begin{align}
-\frac{\hbar^2\nabla^2}{2m}\psi(\mathbf{r})+g\delta^3(\mathbf{r})\frac{\partial }{\partial |\mathbf{r}|}(|\mathbf{r}|\psi(\mathbf{r}))=E\psi(\mathbf{r}).
    	\label{1}
\end{align}
For $\mathbf{r} \neq \mathbf{0}$, up to a normalization factor, wave function $\psi$ is of the form
\begin{align}
\psi(\mathbf{r}) = \frac{sin(qr + \chi)}{r}  +\sum_{l\geq1}\sum_{m=-l}^{m=l} A_{lm}j_l(qr)Y_{lm}(\Omega),
\end{align}
where $\chi$ is the s-wave scattering phase shift, $r=|\textbf{r}|$, $j_l(x)$ is the
spherical Bessel function which is regular at origin [$j_l(x) \sim x^l$
as $x\rightarrow0$ actually], and $Y_{lm}$ is the spherical
harmonic function. Note that for $l \geq 1$, the spherical
Neumann functions $n_l(kr)$ are absent as they are not
square-integrable at $r = 0$ [$n_l(x) \sim x^{-l-1}$ as $\rightarrow 0$].
Consequently, the phase shifts in these channels are also
vanishing. In contrast, for $l = 0$, i.e., the s-wave channel,
both $j_0(r) = sin(r)/r$ and $n_0(r)=cos(r)/r$ are normalizable at $r = 0$ and hence we have to include them both.

Substituting Eq.(6) into Eq.(5), using the well-known formula
\begin{align}
\nabla^2(1/r) =-4\pi\delta^3(\mathbf{r}),
\end{align}
 two delta function terms appear in the left-hand side of Eq.(5). Furthermore, the two delta
functions should cancel each other, so we obtain an implicit equation for the phase shift $\chi$,
\begin{align}
g =-\frac{4\pi\hbar^2}{2mq}\frac{sin(\chi)}{ cos (\chi)}.
\end{align}
In the low energy limit of $q\rightarrow0$, for shorted-ranged potential, the  phase $\chi$ is linear
in $q$, i.e., $\chi\rightarrow -qa_{3D}$ \cite{Landau}.
 We thus get the relation
\begin{align}
g =\frac{4\pi \hbar^2a_{3D}}{2m},
\end{align}
where $a_{3D}$ is the s-wave
scattering length.
Now we see that the effect of the contact pseudo-potential is to generate scattering in the s-wave channel,
while leaving higher partial waves intact.
This is due to the presence of  the centrifugal potential barriers for the higher partial waves, the particle can hardly
feel the short-ranged potential, let alone be scattered. In contrast, the
s-wave wave function can take nonzero values at the origin, so they can
 be scattered.

The above analysis also indicates that the pseudo-potential can be also understood as a boundary condition
on the wave function. For example, by Eq.(6), we know around the contact potential, the wave function is of the form
\begin{align}
\psi(\mathbf{r}) = \frac{c_{-1} }{r} + c_0 + other\ higher \ order \ terms \ in\ r.
\end{align}
The presence of the pseudo-potential is equivalent to the
condition that
\begin{align}
c_0 /c_{-1} = -1/a_{3D}.
\end{align}
In the above equation,  we have used  Eqs.(6), (8) and (9).
So when $r\rightarrow0$, we get the boundary condition of wave function near the origin, i.e. \cite{Demkov},
 \begin{align}
 \label{12p}
\lim_{r\rightarrow0}\psi(\mathbf{r}) \propto 1 /r -\frac{1}{a_{3D}}.
\end{align}


In a three dimension periodic potential, the energy bands usually
need large-scaled numerical calculations.
 In this work, we investigate a rather simple model, which consists of  a three dimension lattice of Huang-Yang pseudopotential.
On one hand, in comparison with one dimensional Kronig-Penney model,  this three dimensional model is much more realistic to simulate the solid state physics.
On the other hand, this model is also simple enough  to deal with mathematically.
\textbf{Due to the relative simplicity of contact potential, the various applications of zero-range (contact) potential  in different situations, e.g., one-center problem, two-center problem, many-center problem, multiple scattering etc are discussed intensively by Demkov and Ostrovskii \cite{Demkov}.}
Furthermore, we also note that the three dimension periodic contact potential model has been used to investigate the neutron diffraction and refraction in solid \cite{Goldberger1947}.
Some properties of the energy bands, i.e., the effective mass of lowest band, the shape of equi-energy surface., have also been reported in previous literatures \cite{Demkov}.

Our work would provide an additional complement to the above literatures on such a model.
 In this work, we find that 
in comparison with single point potential, the presence of  lattice potential changes the existence condition of negative energy states near zero energy.
 It is found that the system always has negative energy states for an arbitrarily weak periodic potential.
In addition, we find that there exists dark states in the energy bands, that some energy bands are not affected by the lattice potential. The physical mechanism of the dark states is explained by  explicitly constructing the standing wave-type Bloch wave functions.

The work is organized as follows. In Sec.\textbf{II}, the model Hamiltonian is given. 
Next, we solve eigenequation for energy bands in Sec.\textbf{III}. The dark states are discussed in Sec.\textbf{IV}.
 At the end, a summary is given in Sec.\textbf{V}.

\section{The model Hamiltonian}
In this work, we consider a Hamiltonian
 \begin{align}
&H=H_0+V_{Lattice}(\mathbf{r})\notag\\
&H_0=\frac{p^2}{2m}=\frac{-\hbar^2\nabla^2}{2m}\notag\\
&V_{Lattice}(\mathbf{r})=\sum_\mathbf{n} g\delta^3(\mathbf{r}-\mathbf{R}_n)\frac{\partial }{\partial |\mathbf{r}-\mathbf{R}_n|}(|\mathbf{r}-\mathbf{R}_n|.),
    	\label{2}
\end{align}
where $V_{Lattice}(\mathbf{r})$ is lattice Huang-Yang psedopotential, $H_0$ is the free particle  Hamiltonian, and $\mathbf{R}_n$ is position vector of pseudo-potential lattice sites.
 Lattice vector $\mathbf{R}_\mathbf{n}=d[n_x,n_y,n_z]$, where $d$ is lattice constant and $n_{i=x,y,z}$ are integers. The reciprocal lattice vector is given by $\mathbf{p}_\textbf{m}=\frac{2\pi}{d}[m_x,m_y,m_z]$ and $m_{i=x,y,z}$ are integers.
 In the whole manuscript, we set $2m=\hbar=d=1$. In such a unit system, the energy is measured by $\frac{\hbar^2}{2m d^2}$, the wave vector is measured by $1/d$, length is measured by $d$ and momentum is measured by $\hbar/d$, respectively.
The dimensionless form of the above Hamiltonian is
 \begin{align}
&H=H_0+V_{Lattice}(\mathbf{r})\notag\\
&H_0=-\nabla^2\notag\\
&V_{Lattice}(\mathbf{r})=g\sum_\mathbf{n}  \delta^3(\mathbf{r}-\mathbf{R}_\mathbf{n})\frac{\partial }{\partial |\mathbf{r}-\mathbf{R}_\mathbf{n}|}(|\mathbf{r}-\mathbf{R}_\mathbf{n}|.),
    	\label{3}
\end{align}
where $g=4\pi a_{3D}$ and $\mathbf{R}_n=[n_x,n_y,n_z]$.

\section{energy bands }
In a periodic lattice potential, the wave function is determined by a homogenous Lippmann-Schwinger equation, i.e,
 \begin{align}
&\psi(\mathbf{r})=\int d^3\mathbf{r'}G^{+}_{0}(\mathbf{r},\mathbf{r}')V_{Lattice}(\mathbf{r}')\psi(\mathbf{r'})\notag\\
&=-\frac{g}{4\pi}\sum_\mathbf{n} \frac{e^{iq|\mathbf{r}-\mathbf{R}_n|}}{|\mathbf{r}-\mathbf{R}_n|}a_\mathbf{n},
\label{201}
\end{align}
where
\begin{align}
G^{+}_{0}(\mathbf{r},\mathbf{r}')=\langle \mathbf{r}|\frac{1}{q^2+i0^+-H_0}|\mathbf{r}' \rangle=\frac{-1}{4\pi}\frac{e^{iq|\mathbf{r}-\mathbf{r}'|}}{|\mathbf{r}-\mathbf{r}'|}
\label{5}
\end{align}
is  Green's function of  free particle Hamiltonian $H_0$ \cite{Economou2006}, and coefficient
\begin{align}
 & a_\mathbf{n}=lim_{\mathbf{r}\rightarrow \mathbf{R}_n}\frac{\partial }{\partial |\mathbf{r}-\mathbf{R}_n|}(|\mathbf{r}-\mathbf{R}_n|\psi (\mathbf{r})).
 \label{701}
\end{align}
 The Bloch's theorem, i.e., $\psi_\mathbf{k}(\mathbf{r+R}_\mathbf{m})=e^{i\mathbf{k}\cdot\mathbf{R}_\mathbf{m}}\psi_\mathbf{k}(\mathbf{r})$, requires $a_\mathbf{n}\propto e^{i\mathbf{k}\cdot\mathbf{R}_\mathbf{n}}$, where $\mathbf{k}$ is a quasi-momentum in the  first Brillouin Zone.
 So the wave function should be (up to a constant factor)
\begin{align}\label{16}
&\psi_\mathbf{k}(\mathbf{r})= \sum_\mathbf{n} \frac{e^{iq|\mathbf{r}-\mathbf{R}_n|}}{|\mathbf{r}-\mathbf{R}_n|}e^{i\mathbf{k}\cdot\mathbf{R}_n}.
\end{align}
As $r\rightarrow 0$, we see that Eq.(\ref{16}) becomes
\begin{align}\label{161}
&\psi_\mathbf{k}(\mathbf{r})= 1/r+\ a \ finite \ number.
\end{align}
Further comparing it with the boundary condition of wave function, i.e., Eq.(\ref{12p}), we get the eigenequation for energy bands
 \begin{align}\label{161}
&-\frac{1}{a_{3D}}=\lim_{r\rightarrow0}[\psi_\mathbf{k}(\mathbf{r})-1/r]\notag\\
&=\lim_{r\rightarrow0}[\sum_\mathbf{n} \frac{e^{iq|\mathbf{r}-\mathbf{R}_n|}}{|\mathbf{r}-\mathbf{R}_n|}e^{i\mathbf{k}\cdot\mathbf{R}_n}-1/r],
\end{align}
where eigenenergy $E=q^2$.
In the followings, we would solve Eq.(\ref{161}) to get the energy bands.

\begin{figure}
\begin{center}
\includegraphics[width=1.1\columnwidth]{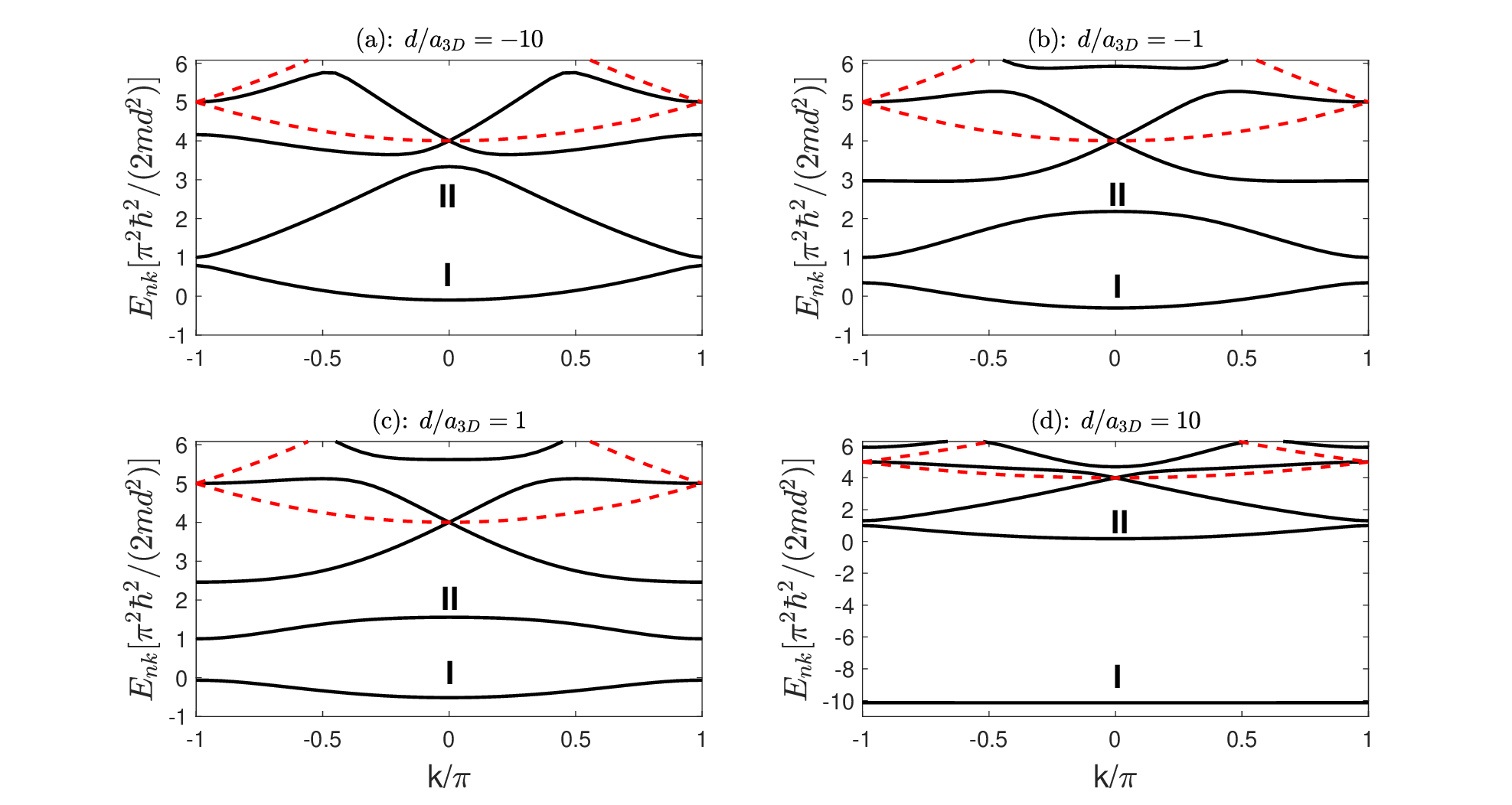}
\end{center}
\caption{ Energy bands of $\mathbf{k}=[0,k_y=k,0]$ for several different potential strengths $d/a_{3D}=-10,-1,1,10$, where $-\pi\leq k\leq\pi$. With the increasing of potential strength $d/a_{3D}$, the lowest band (\textbf{I}) moves down gradually  and the states become negative energy states (energy $E<0$).  The red dashed lines are the energy bands which are not affected by periodic potential.}
\label{Fig3}
\end{figure}


\subsection{summation of a series }
If energy is negative, i.e., $E=q^2<0$, $q$ would be purely imaginary, i.e., $q=i\lambda$ with $\lambda>0$. Then the series
\begin{align}
\pi(\mathbf{r})\equiv\psi_\mathbf{k}(\mathbf{r})=\sum_{n} \frac{e^{iq|\mathbf{r}-\mathbf{R}_n|}}{|\mathbf{r}-\mathbf{R}_n|}e^{i\mathbf{k}\cdot \mathbf{R}_n}
\label{19}
\end{align}
in the Eq.(\ref{161}) is absolutely convergent. While when energy is positive, $q$ can  be taken as positive real number. In such a case,  the series in (\ref{19}) is conditionally convergent.
The finial result usually depends on the order of summation.
  In order to get a definite number,  in the following, we would regularize the series by the Abel method  and transform it into an absolutely convergent series.
 Namely, we shall imagine that $q$ has a positive imaginary part $\epsilon$, so that the summation is absolutely convergent and has a well-defined value. At the end, we then let $\epsilon\rightarrow 0 $, and hopefully the summation has a limiting value, and then the limiting value will be assigned to the original summation.

In order to complete the summation of  series,
we would use Ewald's summation method \cite{Ewald1921,Borwein2013,Born1954}.
The following integral identity
\begin{eqnarray}
\label{Identity}
\frac{e^{-\alpha R}}{R} &=& \frac{2}{\sqrt{\pi}} \int_0^\infty e^{-R^2 t^2 - \alpha^2/4t^2}  d t ,\ \alpha > 0 , R>0 .
\end{eqnarray}
is of use.
The proof is in the \textbf{Appendix A}. But what we really want is an integral representation of the factor $e^{iq R}/R$ [see Eq.(\ref{19})]. The integral diverges at the origin for $\alpha =  -i q $. We thus have to deform the integration path so that the continuation $\alpha \rightarrow -i q $ is legitimate. For this purpose, we can take the path that leaves the origin in the direction $\text{Arg } t = \text{Arg } q - \pi/4$ and then return to the real axis \cite{Borwein2013}. Denoting the path by $\mathcal{C}$, we have
\begin{eqnarray}
  \frac{e^{i q  R}}{R} &=& \frac{2}{\sqrt{\pi}} \int_\mathcal{C} e^{-R^2 t^2 + q^2/4t^2}  d t.
\end{eqnarray}
With this integral representation, we have
\begin{eqnarray}\label{defpsi}
  \pi(\mathbf{r}) &=&  \frac{2}{\sqrt{\pi}} \int_\mathcal{C} \sum_{\textbf{n}} e^{\left[ -(\textbf{R}_\textbf{n} - \textbf{r})^2 t^2 + i \textbf{k}\cdot \textbf{R}_\textbf{n} + \frac{q^2}{4t^2} \right]} dt .
\end{eqnarray}
We have a periodic sampling of the function
\begin{eqnarray}
 g_\textbf{r} (\textbf{R}) =  e^{\left[ -(\textbf{R} - \textbf{r})^2 t^2 + i \textbf{k}\cdot \textbf{R} \right]} .
\end{eqnarray}
The summation is thus amenable to the Poisson summation formula. Its Fourier transform is
\begin{eqnarray}
 && \hat{g}_\textbf{r} (\textbf{p}) = \iiint d^3\textbf{R}  e^{\left[ -(\textbf{R} - \textbf{r})^2 t^2 + i \textbf{k}\cdot \textbf{R} -i \textbf{p} \cdot \textbf{R}\right]}  \nonumber \\
  &&=  \frac{\pi^{3/2}}{t^3 } e^{  -\frac{(\textbf{k}-\textbf{p})^2+i \textbf{r}\cdot (\textbf{k}-\textbf{p})}{4t^2} } .
\end{eqnarray}
By the Poisson summation formula, we know
\begin{eqnarray}
  \sum_{\textbf{n}}  g_\textbf{r} (\textbf{R}_\textbf{n}) &=&   \sum_{\textbf{m}}  \hat{g}_\textbf{r} (\textbf{p}_\textbf{m}),
\end{eqnarray}
where on the right-hand side, the summation is over all reciprocal lattice sites, $\textbf{p}_\textbf{m} = 2\pi \textbf{m} $.
So we have
\begin{eqnarray}\label{defpsi1}
  \pi (\mathbf{r}) &=& 2\pi  \int_\mathcal{C} \sum_{\textbf{m}} e^{  -\frac{ (\textbf{k}-\textbf{p}_\textbf{m})^2 - q^2}{4t^2} + i \textbf{r}\cdot (\textbf{k}-\textbf{p}_\textbf{m}) } \frac{dt}{t^3} .
\end{eqnarray}
Now there are two equivalent expressions for $\pi (\mathbf{r})  $, involving two equivalent integrands. One is a summation over the real lattice sites Eq.(\ref{defpsi}), while the other is a summation over reciprocal lattice sites Eq.(\ref{defpsi1}). Next we divide the range of the $t$-integration at $\eta$ (assumed to be real) and in each segment we perform the integration with a different summation. We have thus
\begin{eqnarray}
 \pi (\mathbf{r}) &=& \pi^{(1)} (\mathbf{r}) +\pi^{(2)} (\mathbf{r}), \nonumber \\
  \pi^{(1)} (\mathbf{r}) &=& 2\pi \int_{(0)}^\eta \sum_{\textbf{m}} e^{  -\frac{ (\textbf{k}-\textbf{p}_\textbf{m})^2 - q^2}{4t^2} + i \textbf{r}\cdot (\textbf{k}-\textbf{p}_\textbf{m}) } \frac{dt}{t^3} ,\nonumber \\
  \pi^{(2)} (\mathbf{r}) &=& \frac{2}{\sqrt{\pi}} \int_\eta^\infty  \sum_{\textbf{n}}e^{ -(\textbf{R}_\textbf{n} - \textbf{r})^2 t^2 + i \textbf{k}\cdot \textbf{R}_\textbf{n} + \frac{q^2}{4t^2} } dt .
\end{eqnarray}
The integral in $ \pi^{(1)} (\mathbf{r}) $ is elementary, while the integral in $ \pi^{(2)} (\mathbf{r}) $ leads to the error function. In the \textbf{Appendix B}, we give the detailed calculation. In the end, we have
\begin{eqnarray}\label{error}
  &&\pi^{(1)} (\mathbf{r})
  = 4\pi \sum_{\textbf{m}} \frac{ e^{-\frac{[(\textbf{k}-\textbf{p}_\textbf{m})^2 - q^2]}{4 \eta^2} + i\textbf{r}\cdot (\textbf{k}-\textbf{p}_\textbf{m}) }}{(\textbf{k}-\textbf{p}_\textbf{m})^2 - q^2} , \notag\\
 && \pi^{(2)} (\mathbf{r})= \sum_{\textbf{n}}  \frac{e^{i \textbf{k}\cdot \textbf{R}_\textbf{n}}}{|\textbf{R}_\textbf{n}-\textbf{r}|} \frac{1}{2}\bigg[ e^{i q |\textbf{R}_\textbf{n}-\textbf{r}| }  erfc(|\textbf{R}_\textbf{n}-\textbf{r}| \eta + \frac{i q \eta}{2} )\notag\\
&& \quad  +e^{-i q |\textbf{R}_\textbf{n}-\textbf{r}| } erfc(|\textbf{R}_\textbf{n}-\textbf{r}| \eta - \frac{i q}{2 \eta} ) \bigg] .
\end{eqnarray}
where the free parameter $\eta>0$ is used to control the convergent speed of the series and $erfc(x)=1-erf(x)$ is the complementary error function, and the error function is $erf(x)=\frac{2}{\sqrt{\pi}}\int_{0}^{x} e^{-t^2}dt$ [see \textbf{Appendix B}]. In order to get the expression for $\pi^{(1)}$ in the above equation, we used the Abel's regularization method.

We now need to determine the asymptotic behavior of the wave function $\pi(\mathbf{r})$. First, we note that $\pi^{(1)}$ is regular at $\mathbf{r} = 0$. Second, it is easy to verify that the divergent part of $\pi^{(2)}$ is $1/r$. Actually, it comes form the $\textbf{n} =0 $ term. The constant part of $\pi^{(2)}$ is then
\begin{eqnarray}
 & &\lim_{r\rightarrow 0 }\left[ \pi^{(2)}(\mathbf{r}) -\frac{1}{r} \right ]\notag\\
  & &=   \sum_{\textbf{n} \neq 0 }  \frac{e^{i \textbf{k}\cdot \textbf{R}_\textbf{n}}}{|\textbf{R}_\textbf{n}|} \frac{1}{2}\bigg[ e^{i q |\textbf{R}_\textbf{n}| } erfc(|\textbf{R}_\textbf{n}| \eta + \frac{i q}{2 \eta} )\nonumber \\
  &&  \quad \quad \quad \quad \quad \quad \quad \quad +e^{-i q |\textbf{R}_\textbf{n}| } erfc(|\textbf{R}_\textbf{n}| \eta - \frac{i q}{2 \eta} ) \bigg]  \nonumber \\
 & &  +  \lim_{r\rightarrow 0 } \left\{\frac{1}{2r} [e^{iq r } erfc(\eta r + \frac{iq}{2\eta} ) +e^{-iq r } erfc(\eta r - \frac{iq }{2\eta} )] -\frac{1}{r} \right\} \nonumber \\
  &&=   \sum_{\textbf{n} \neq 0 }  \frac{e^{i \textbf{k}\cdot \textbf{R}_\textbf{n}}}{|\textbf{R}_\textbf{n}|} \frac{1}{2}\bigg[ e^{i q |\textbf{R}_\textbf{n}| } erfc(|\textbf{R}_\textbf{n}| \eta + \frac{i q}{2 \eta} )\nonumber \\
  && \quad \quad \quad \quad \quad \quad \quad \quad +e^{-i q |\textbf{R}_\textbf{n}| } erfc(|\textbf{R}_\textbf{n}| \eta - \frac{i q}{2 \eta} ) \bigg]  \nonumber \\
  &&  \quad \quad \quad \quad \quad \quad \quad \quad-iq erf(\frac{i q}{2 \eta}) - \frac{2}{\sqrt{\pi}} \eta e^{\frac{q^2}{4\eta^2} }.
\end{eqnarray}
The eigenequation is then
\begin{eqnarray}
\label{321}
  -\frac{1}{a_{3D}} &=&  \lim_{r\rightarrow 0 } [ \pi(\textbf{r}) -\frac{1}{r} ]\nonumber \\
             &=&\pi^{(1)} (0) + \lim_{r\rightarrow 0 } \left[ \pi^{(2)}(\mathbf{r}) -\frac{1}{r} \right ] .
\end{eqnarray}

  In our following numerical calculations (with Mathematica soft ware's help), we take parameter $\eta=3$ and first $7\times7\times7=343$ terms in $\pi^{(1)}$ and  $\pi^{(2)}$ to approximate the above series.
\textbf{And then, for a fixed quasi-momentum $\textbf{k}$ of Brillouin Zone, we use the command FindRoot to search the root (eigenenergy) of Eq.(\ref{321}). The over all error of energy bands is about $10^ {-6}$. }

\subsection{Results }
\begin{figure}
\begin{center}
\includegraphics[width=1.0\columnwidth]{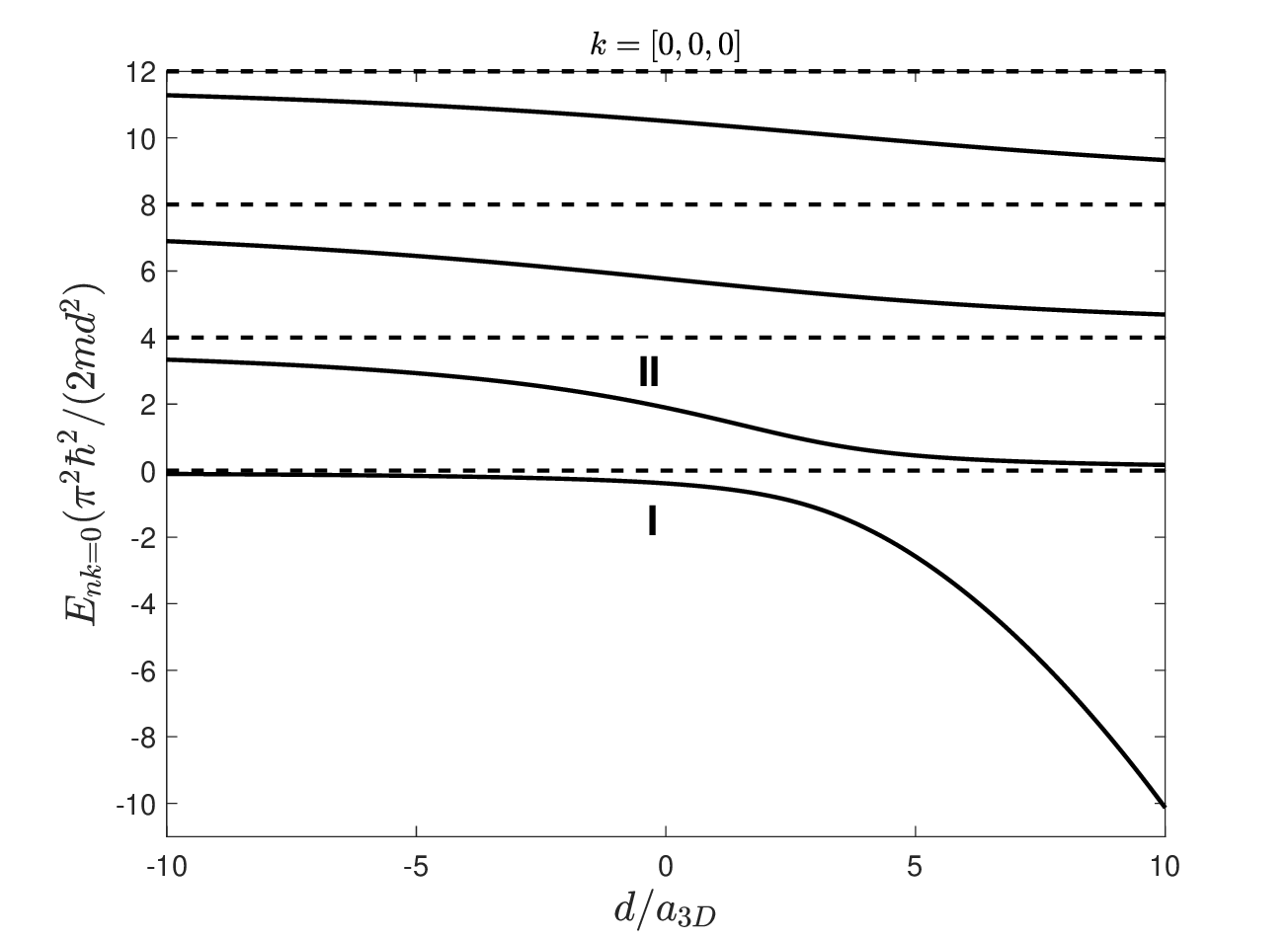}
\end{center}
\caption{ Eigenenergy of $\mathbf{k}=[0,0,0]$ as a function of  potential strength $d/a_{3D}$. The dashed lines are energies of free particle (potential strength $g=0$).  When $d/a_{3D}\rightarrow+\infty$, the lowest energy of band \textbf{II} would be $E_{\mathbf{II},\mathbf{k}=[0,0,0]}=4\pi a_{3D}\rightarrow0^+$ [see Eq. (\ref{261})].}
\label{Fig4}
\end{figure}

We report the energy bands of $\mathbf{k}=[0,k_y=k,0,]$ in Fig.\ref{Fig3}  for different potential strengths $d/a_{3D}=-10,-1,1,10$. We see that the system has a lot of energy bands due to the periodic potential.
With increasing of $d/a_{3D}$, a lot of the energy bands  are pulled down gradually (see the black solid  lines ).
When the potential strength is weak , e.g., $d/a_{3D}=-10$, the lowest energy band (\textbf{I}) moves down a bit. When the potential is very strong,  e.g., $d/a_{3D}=10$, the lowest every band (\textbf{I}) is shifted downward substantially (see also Fig.\ref{Fig4}).

\begin{figure}
\begin{center}
\includegraphics[width=1.0\columnwidth]{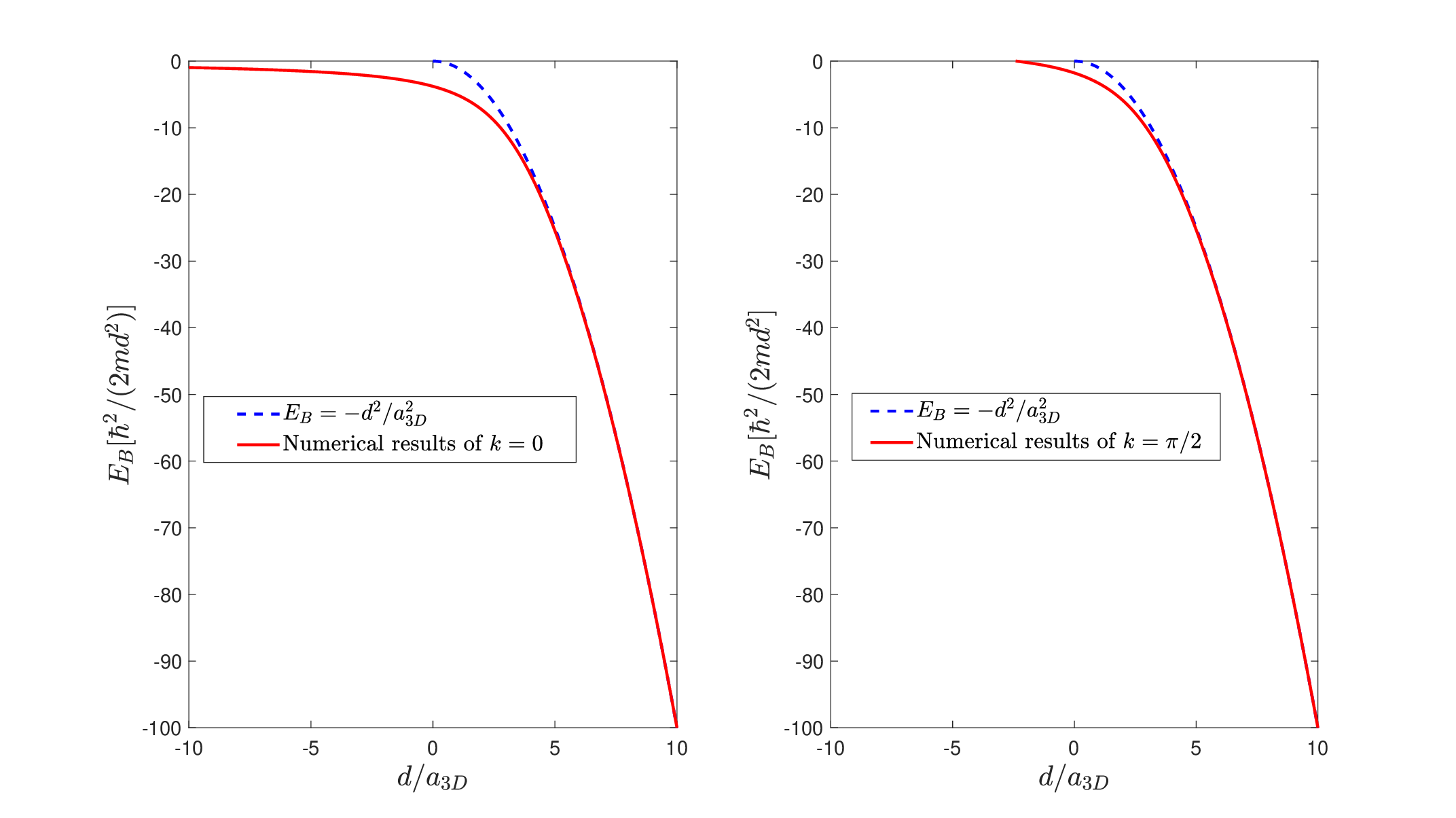}
\end{center}
\caption{ The negative energy part of the first energy band ($E<0$) of $\mathbf{k}=[0,k,0]$, where $k=0$ for panel (a) and $k=\pi/2$ for panel (b). When $d/a_{3D}\rightarrow-\infty$, the lowest energy of band \textbf{I} would be $E_{\mathbf{I},\mathbf{k}=[0,0,0]}=4\pi a_{3D}/d^3\rightarrow0^-$ [see Eq. (\ref{26})].
 While when $d/a_{3D}\rightarrow\infty$, the eigenenergy would be the bound state energy of single point potential ($E_{\mathbf{I},\mathbf{k}=[0,0,0]}\simeq E_B=-1/a_{3D}^{2}$). }
\label{Fig5}
\end{figure}

Fig.\ref{Fig4} gives the evolution of eigenenergies of $\mathbf{k}=[0,0,0]$ with the increasing of potential strength  $d/a_{3D}$.
The dashed lines are the energies  of system when the lattice potential is turned off, i.e., potential  strength $g=0$.
We see that with the increasing of potential strength $d/a_{3D}$, the lowest band labeled with \textbf{I} moves down without limits.
In fact, when the potential strength  $d/a_{3D}$ is a very large positive number, the energy of band \textbf{I} would  be approximately the bound state energy of single point potential, i.e., $E_{\mathbf{I},\mathbf{k}=[0,0,0]}\rightarrow E_B=-1/a_{3D}^2$ as $d/a_{3D}\rightarrow\infty$ (see Fig.\ref{Fig5}).

For other bands. e.g., band \textbf{II}, the eigenenergies would approach some free particle energies ($g=0$) (see dashed lines of Fig.\ref{Fig4}].
We see that as the potential strength $d/a_{3D}\rightarrow+\infty$, the lowest point of band \textbf{II} approaches zero which is the lowest energy of free particle.
These results show that as the potential strength $d/a_{3D}\rightarrow+\infty$ ($a_{3D}\rightarrow0^+$), for lowest band \textbf{I}, the potential is strongly attractive, the states of energy band \textbf{I} are the superpositions of the  tightly bound states.
While for band \textbf{II}, if the energy band \textbf{I} is ignored intentionally, the states in the band \textbf{II} can be viewed as states of lowest band which experience weak repulsive potential.
In this sense, the positive scattering length corresponds a repulsive potential.

\textbf{In other word, starting from free particle limit, when one turns scattering $a_{3D}$ from $0^-$ to $-\infty$, the particle experiences an attractive potential.
 While when $a_{3D}$ goes from $0^+$ to $+\infty$, the particle would feel a repulsive potential.
The above similar multiple energy band physics have also been found in the BCS-BEC crossover of fermion atom gas \cite{Zwerger}, in which when $1/a_{3D}$ goes from $-\infty$ to $+\infty$, the system has two distinct energy branches. The low branch energy corresponds to the stable ground state, the upper branch is the metastable exited state.
}

Fig.\ref{Fig5} gives the negative energy part of first band energy band ($E<0$) for two different quasi-momenta $\mathbf{k}=[0,0,0]$ and $\mathbf{k}=[0,\pi/2,0]$.
\textbf{It is well known that for an isolated single point potential , only when the s-wave scattering length is positive, i.e., $d/a_{3D}>0$, the system has  a bound state with a negative energy $E_B=-1/a^{2}_{3D}$ \cite{Ueda} (see the blue dashed lines of Fig.\ref{Fig5}).}
From the panel (a) of Fig. \ref{Fig5}, we see that in the presence of lattice potential, even when potential is very weak, i.e, $d/a_{3D}\rightarrow -\infty$ ($a_{3D}\rightarrow0^-$), the system also has negative energy states near zero energy for $\mathbf{k}=[0,0,0]$. The existence conditions of negative energy states have been changed.
This is because when $d/a_{3D}\rightarrow-\infty$ and $\mathbf{k}=[0,0,0]$, the left-hand side of Eq.(\ref{321}) is a very large positive number.
 While  the right-hand side of Eq.(\ref{321}) is basically given by $\pi^{(1)}(0)$ [see Eq.(\ref{error})], which can be approximated by the single term $\mathbf{m}=\mathbf{0}$ ($\mathbf{p}_m=0$), i.e.,
 \begin{align}\label{25}
& \pi^{(1)}(\mathbf{r=0})=4\pi\sum_{\mathbf{m}}\frac{e^{-\frac{(\mathbf{k}-\textbf{p}_m)^2-q^2}{4\eta^2}}}{(\mathbf{k}-\textbf{p}_m)^2-q^2}\notag\\
&=\frac{4\pi}{-q^2} +\ some \ finite \ number\notag\\
&\sim\frac{4\pi}{-q^2}\rightarrow +\infty
\end{align}
as $q^2\rightarrow0^-$ (for negative energy states). Then based on Eq. (\ref{321}), the negative energy is approximately
\begin{align}\label{26}
E_{\mathbf{I},\mathbf{k}=[0,0,0]}=q^2\simeq 4\pi a_{3D}<0
\end{align}
as $d/a_{3D}\rightarrow-\infty$ ($a_{3D}\rightarrow 0^-$).
So the negative energy states always exist for arbitrarily weak potential, i.e., $d/a_{3D}\rightarrow -\infty$.
Panel (b) shows that the existence region of negative energy states shrinks when quasi-moment $k$ increases from $0$ to $\pi$.
For a finite $k\neq0$, because the right-hand side of Eq.(\ref{19}) becomes finite now, only when $d/a_{3D}$ is sufficiently large, the system has negative energy states.
This is reflected the fact that the factor $e^{i\mathbf{k}\cdot\mathbf{R}_n}$ has phase destructive interference effects in the series of Eq. (\ref{19}).

When $\mathbf{k}=[0,0,0]$ and $d/a_{3D}\rightarrow +\infty$ ($a_{3D}\rightarrow 0^+$), the lowest energy of band \textbf{II} (see Fig.\ref{Fig4}) can be similarly given  by
\begin{align}\label{261}
E_{\mathbf{II},\mathbf{k}=[0,0,0]}=q^2\simeq 4\pi a_{3D}>0.
\end{align}
The positiveness of the energy indicates that when $a_{3D}>0$, it seems that the particles  experience an effective repulsive potential.

\begin{table}
\caption{The four high symmetrical points in the first Brillouin Zone \cite{Setyawan2010}. Here vectors $\mathbf{b}_1=\frac{2\pi}{d}[1,0,0]$, $\mathbf{b}_2=\frac{2\pi}{d}[0,1,0]$ and $\mathbf{b}_3=\frac{2\pi}{d}[0,0,1]$.
}
\begin{center}
\begin{tabular}{|c|c|c|c|c|c|c|c|c|c|c|}
\hline
\hline
\hline
$\mathbf{b}_1$&$\mathbf{b}_2$&$\mathbf{b}_3$& & $\mathbf{b}_1$&$\mathbf{b}_2$&$\mathbf{b}_3$&\tabularnewline
\hline
 0&0& 0 &  $\mathbf{\Gamma}$  &1/2& 1/2&0 & $\mathbf{M}$ \tabularnewline
 \hline
 0&1/2& 0 & $\mathbf{X}$  & 1/2& 1/2&1/2 & $\mathbf{R}$  \tabularnewline
 \hline
\end{tabular}
\end{center}
\end{table}
\begin{figure}
\begin{center}
\includegraphics[width=1.00\columnwidth]{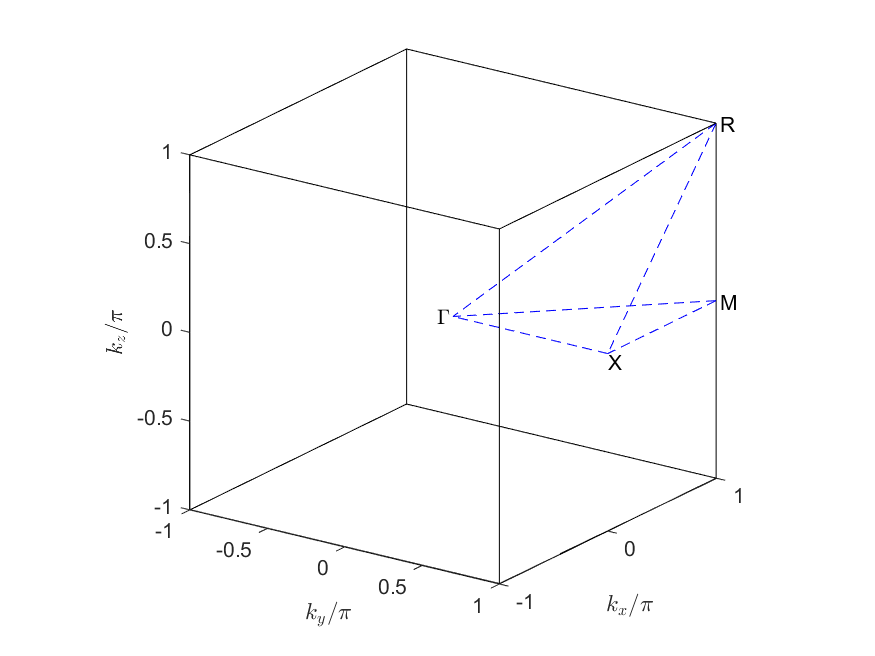}
\end{center}
\caption{ A path connecting the  four high symmetrical points in the first Brillouin Zone }
\label{Fig61}
\end{figure}

\begin{figure}
\begin{center}
\includegraphics[width=1.0\columnwidth]{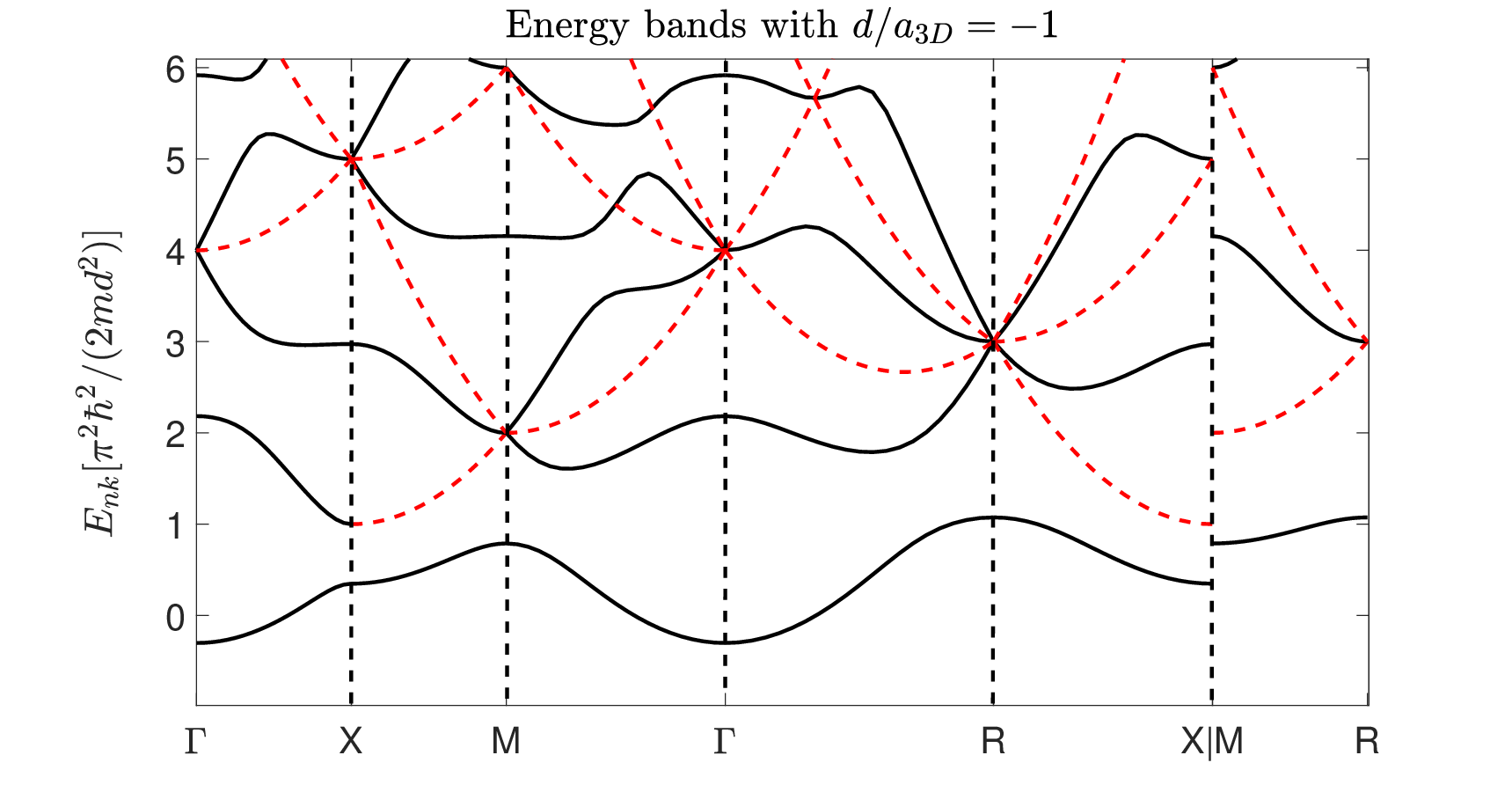}
\end{center}
\caption{ Energy bands of the first Brillouin Zone  with $d/a_{3D}=-1$. The red dashed lines are the energy bands which are not affected by periodic potential. Their eigenstates are called as dark states. }
\label{Fig6}
\end{figure}

The four high symmetric points $\mathbf{\Gamma}$, $\mathbf{X}$, $\mathbf{M}$, $\mathbf{R}$ are defined in Table.\textbf{I}, which is consistent with Ref. \cite{Setyawan2010}.
Along a path in the first Brillouin Zone (see Fig.\ref{Fig61}), we report the energy bands in Fig. \ref{Fig6}
for a specific potential strength $d/a_{3D}=-1$.
The red dashed lines are the energy bands which are not affected by the periodic potential.
It is found that the eigenenergies of black solid lines are usually non-degenerate, while the red lines can be multiple degenerate.

The appearances of the red dashed lines and degeneracy of energy bands can be explained with the group theory. Here for the simple cubic lattice, the point group of  lattice is the $O_h$ group.
For a given wave vector $\mathbf{k}$, there may exist a subgroup of $O_h$, that leaves the wave vector $\mathbf{k}$ unchanged, i.e., the group of wave vector $G(\mathbf{k})$.
For example, for a wave vector $\mathbf{k}=[k,0,0]$  ($0<k<\pi$), its group $G(\mathbf{k})$ is $C_{4v}$ which has $x-$ axis of fourth order  \cite{Dresselhaus}.
We know that the wave functions of energy bands at $\mathbf{k}$ usually form the basis functions of an irreducible representation of the group $G(\mathbf{k})$.
 For a general point $\mathbf{k}$ in first Brillouin Zone, due to the triviality of the group of wave vector, i.e., the unique group element is the identity element, its irreducible representation is the one dimension trivial representation, then the energies at $\mathbf{k}$ are non-degenerate. In such a case, generally speaking, the energy bands can be affected by the periodic potential.

When wave vector $\textbf{k}$ is a symmetric point in first Brillouin Zone, e.g., all the $\mathbf{k}$ points given by Fig.\ref{Fig6} ,  the group $G(\mathbf{k})$ is not trivial, then it can have non-trivial irreducible representations.
In the \textbf{Appendix C}, we prove that if an irreducible representation is not a trivial representation of the group of wave vector $G(\mathbf{k})$, its basis functions at origin would be zero.
Then due to the periodicity of the lattice, the wave function is vanishing at all the lattice sites.
In addition, because the Huang-Yang pseudopotential is short ranged,
the Huang-Yang pseudopotential would have no effect on the wave functions.
  Consequently the corresponding energy band also could not be changed (relative to the energy of free particle)(see the red dashed lines in Fig.\ref{Fig6}).
Moreover,  the dimension of a non-trivial irreducible representation can be larger than one,  then the energy bands of red dashed lines can be multiple degenerate (see also the discussions of next section).

When a wave function at origin takes a finite value, generally the corresponding energy band would be changed  (relative to the energy of free particle)(the black solid lines of Fig.\ref{Fig6}).
  This is because the wave functions of black lines would be basis functions of  trivial representations of the group of wave vector.
 Furthermore, because all the trivial irreducible representations are always one dimensional, then the corresponding energy bands are in fact non-degenerate.
The above discussions indicate that only when the wave functions of  energy bands form  a trivial  representation of the group of wave vector, the energy band can be shifted by the periodic potential.

\section{Dark states}
From the red dashed lines of Figs. \ref{Fig3}, and \ref{Fig6}, we see there are some energies which are not  affected by the lattice potential. Accordingly, we would call the corresponding eigenstates as dark stats.  For example, in the band \textbf{II} of Fig.\ref{Fig3}, when the quasi-momentum (wave vector) $\mathbf{k}=[0,\pm \pi,0]$,
the energy $E=q^2=\pi^2$ regardless of potential strength $d/a_{3D}$.
In order to further understand such a phenomenon, we start from the one dimensional cases.

In one dimensional lattice [Kronig-Penney model Eq.(\ref{Penny})], the potential is
\begin{align}\label{27}
V_{1D}(x)=\sum_{n=-\infty}^{\infty}g_{1D} \delta(x-n),
\end{align}
where potential strength $g_{1D}$ is measured by $\frac{\hbar^2}{2md}$.
In such a case, similarly the wave function is given by
\begin{align}\label{28}
&\psi_k(x)=\sum_n e^{iq|x-n|}e^{ikn}.
\end{align}
It satisfies Bloch theorem, i.e., $\psi_k(x+m)=e^{ikm}\psi_k(x)$.
 Substituting it into Schr\"{o}dinger equation, i.e.,
\begin{align}\label{29}
[E-H_0]\psi_k(x)=V_{1D}\psi_k(x)=\sum_ng_{1D}\delta(x-n)\psi_k(x),
\end{align}
 and using the fact that $[E+d^2/dx^2]e^{iq|x-n|}=2iq\delta(x-n)$ (see the proof in \textbf{Appendix D}), then we get
 \begin{align}\label{30}
&2iq\sum_n \delta(x-n)e^{ikn} =g_{1D}\sum_{m} \delta(x-m)\psi_k(x)\notag\\
&\int_{0^-}^{0^+}dx[2iq\sum_n \delta(x-n)e^{ikn}] \notag\\
&=\int_{0^-}^{0^+}dx[g_{1D}\sum_{m} \delta(x-m)\psi_k(x)]\notag\\
&2iq=g_{1D}\psi_k(x=0)=g_{1D}\sum_n e^{iq|n|}e^{ikn}\notag\\
&or\rightarrow \frac{-1}{g_{1D}}=\frac{\sum_ni e^{iq|n|}e^{ikn}}{2q}.
\end{align}

In order to further reveal the origin of dark states, we need to transform the above real space summation into momentum space one.
Using formula $\frac{1}{2\pi}\int d k_1\frac{e^{ik_1x}}{k_{1}^{2}-(q^2+i0^+ )}=\frac{ie^{iq|x|}}{2q}$ and Poisson summation formula
$\sum_ne^{ipn}=2\pi\sum_m\delta(p-2\pi m)$,
we  transform Eq.(\ref{30}) into  a summation over the reciprocal lattice site, i.e.,
 \begin{align}\label{311}
&-\frac{1}{g_{1D}}=\frac{\sum_ni e^{iq|n|}e^{ikn}}{2q}\notag\\
&=\frac{1}{2\pi}\int dk_1\sum_n\frac{e^{i(k_1+k)n}}{k_{1}^{2}-(q^2+i0^+ )}\notag\\
&=\frac{2\pi}{2\pi}\int dk_1\sum_m\frac{\delta[(k_1+k)-2\pi m]}{k_{1}^{2}-q^2}\notag\\
&=\sum_m\frac{1}{(2\pi m-k)^{2}-q^2}.
\end{align}
Further using formula $\sum_{m=-\infty}^{+\infty}\frac{1}{2\pi m-z}=-\frac{cos(z/2)}{2sin(z/2)}$, finally the eigenequation is reduced to
\begin{align}\label{31}
-\frac{1}{g_{1D}}=&\sum_m \frac{1}{(2\pi m-k)^2-q^2}\notag\\
=&\frac{1}{2q}\sum_m[\frac{1}{(2\pi m-k)-q}-\frac{1}{(2\pi m-k)+q}]\notag\\
=&\frac{1}{2q}[\frac{-cos[(k+q)/2]}{2sin[(k+q)/2]}-\frac{-cos[(k-q)/2]}{2sin[(k-q)/2]}]\notag\\
=&\frac{sin(q)}{2q[cos(q)-cos(k)]},
\end{align}
which is consistent with Landau-Lifshitz's book \cite{Lifshitz1980}. After adding an infinitesimal imaginary part to wave vector $q$, i.e. $q\rightarrow q+i0^+$,  this final result can be also simply obtained by calculating the geometric series in right-hand side of Eq.(\ref {30}).

When  $k\rightarrow  \pi$,  and $q\rightarrow \pi$, there exist two divergent terms [$m=0$ and $m=1$], i.e.,
\begin{align}\label{32}
&Z\equiv\frac{1}{(2\pi*0-k)^2-q^2}+\frac{1}{(2\pi*1-k)^2-q^2}\notag\\
&=\frac{1}{k^2-q^2}+\frac{1}{(2\pi-k)^2-q^2},
\end{align}
which dominate over other terms in the summation Eq.(\ref{311}).
In fact, one can show that,  depending on how $q$ and $k$ approach $\pi$, the sum of above two terms $Z$ can take any real value from $-\infty$ to $+\infty$.
So the eigenequation Eq.(\ref{311}) can be written as
\begin{align}\label{33}
&-\frac{1}{g_{1D}}=Z+\ some \ finite \ number.
\end{align}
Then the eigenequation can  be always satisfied if $q$ and $k$ take appropriate ways to approach $\pi$.
Consequently, when $k=\pi$, the eigenenergy in the energy band \textbf{II} always takes the free particle energy $E=q^2=\pi^2$ no matter how strong the lattice potential strength is.

 In a similar fashion, one can also analyze the divergent terms of $\pi^{1}(\mathbf{r}=0)$ in Eq.(\ref{error}) to   discuss the dark states in three dimension cases.
Such an effect is due to the divergences of some [at least two (for the reasons please see the following discussions)] terms  in the eigenequation, then the resultant eigenenergies could not be affected by the lattice potential.

The dark states can be also understood as follows. Due to the short-ranged  potential, if the lattice sites are exactly the nodes of wave function, the wave function would not experience the existence of lattice potential .
Then the wave function should satisfy  the free particle Schr\"{o}dinger equation, i.e.,
\begin{align}\label{34}
[E-H_0]\psi_k(\mathbf{r})=[q^2+\nabla^2]\psi_k(\mathbf{r})=0.
\end{align}
Furthermore,  the Bloch wave function can be expanded with plane waves, i.e,
\begin{align}
\psi_k(\mathbf{r})=\sum_{\mathbf{m}}C_{m}e^{-i2\pi \mathbf{m}\cdot \mathbf{r}}e^{i\mathbf{k}\cdot\mathbf{r}}.
\label{35}
\end{align}
where $C_\mathbf{m}$ is the expansion coefficient.
Substituting it in Eq.(\ref{34}), we get
\begin{align}
\sum_{\mathbf{m}}[q^2-(2\pi \mathbf{m}-\mathbf{k})^2]C_{\mathbf{m}}e^{-i2\pi \mathbf{m}\cdot \mathbf{r}}e^{i\mathbf{k}\cdot\mathbf{r}}=0.
\label{36}
\end{align}
Then, for non-vanishing $C_\mathbf{m}$, the energy should be
\begin{align}
&E=q^2=(2\pi \mathbf{m}-\mathbf{k})^2\notag\\
&=(2m_x-k_x)^2+(2m_y-k_y)^2+(2m_z-k_z)^2.
\label{37}
\end{align}
which is free particle energy with a wave vector $2\pi(\mathbf{m}-\mathbf{k})$.

In addition, nodes of wave function at lattice sites imply that the wave function satisfies
\begin{align}
\psi_k(\mathbf{r}=0)=0\Rightarrow \sum_{\mathbf{m}} C_{\mathbf{m}}=0.
\label{38}
\end{align}
One can solve Eqs.(\ref{37}) and (\ref{38}) to get some integer solutions $\mathbf{m}=[m_x,m_y,m_z]$, and then substitute it in  Eq.(\ref{35}) to get the wave function $\psi_k(\mathbf{r})$.
We can see that in order to satisfy Eq.(\ref{38}), the number of (nontrivial) integer solutions ($m_x,m_y,m_z$) to Eq.(\ref{37})  should be two at least.
So the energy $E=q^2$ should correspond to the degenerate points of the folded free particle energy bands in the first Brillouin Zone.

Furthermore, for one dimension cases, one can get two types of solutions. One is $m=n$ or  $m=-n+1$ ($n=0,1,2,3,...$), and $k=\pi, q=(2n-1)\pi$ . The other is  $m=\pm n$ ($n=1,2,3,...$), and $k=0$, $q=2 n\pi$. Further substituting them in Eq.(\ref{35}), we can construct two types of  standing wave-type Block wave functions, i.e,
\begin{align}\label{39}
&\psi_{k=\pi}(x)=[e^{i(2n-2)\pi x}-e^{-i2n\pi x}]e^{i\pi x}\propto sin[(2n-1)\pi x]\notag\\
&\psi_{k=0}(x)=[e^{i2n\pi x}-e^{-i2n\pi x}]e^{i0x}\propto sin(2n\pi x).
\end{align}

For three dimension cases, because the dimension of irreducible representation can be larger than one, the energy $E=q^2$ is usually multiple degenerate.
For example, the lowest red dashed line (energy $4\pi^2\rightarrow 5\pi^2$) of $\mathbf{\Gamma}\rightarrow \mathbf{X}$  in Fig.\ref{Fig6} corresponds the following three wave functions
\begin{align}\label{39}
&(a):\ \mathbf{m}=[2,0,0] \ or  \ \mathbf{m}=[-2,0,0]  \notag\\
& \mathbf{k}=[0,k_y,0], \ E=4\pi^2+k_{y}^2,  \ \psi^{a}_{\mathbf{k}=[0,k_y,0]}=sin(2\pi x) e^{ik_y y}\notag\\
&(b):\ \mathbf{m}=[0,0,2] \ or  \ \mathbf{m}=[0,0,-2]  \notag\\
& \mathbf{k}=[0,k_y,0], \ E=4\pi^2+k_{y}^2,  \ \psi^{b}_{\mathbf{k}=[0,k_y,0]}=sin(2\pi z) e^{ik_y y}\notag\\
&(c):\ \mathbf{m}=[\pm 2,0,0] \ or\ \mathbf{m}=[0,0,\pm 2] \notag\\
&\mathbf{k}=[0,k_y,0], \ E=4\pi^2+k_{y}^2 \notag\\
& \psi^{c}_{\mathbf{k}=[0,k_y,0]}=[cos(2\pi x)-cos(2\pi z) ]e^{ik_y y},
\end{align}
where $0\leq k_y\leq \pi$.
We see that when a dark state takes place,  in general, the wave function is a superposition of finite number plane waves. It indicates that the wave function of a dark state is  much simpler than a general Bloch wave function, which is usually a superposition of infinite number plane waves.

Finally, we should emphasize that although the energy is not affected by the periodic potential, due to restriction of Eq.(\ref{38}) on the wave function, the eigenstate is not a free plane wave state.  This is due to the presence of periodic potential, the momentum is not a good quantum number and a (single) plane wave state is not an eigenstate of full Hamiltonian.

\section{summary}
We investigate  the energy bands of a simple cubic lattice of Huang-Yang pseudopotential.
The energy bands, the existence conditions of negative states, the degeneracy of energy bands are discussed in detail.
In addition, it is found that due to the short ranged potential, there exist dark states in the energy bands, that the wave function is a superposition of finite number plane waves. The physical mechanism of dark states is explained by explicitly constructing standing wave-type Bloch wave functions.
The physics of this model  may be realized in cold atom experiments \cite{Massignan2006}.

\section*{Acknowledgements}
This work was supported by the NSFC under Grants Nos.
11874127, the Joint Fund with
Guangzhou Municipality under No.
202201020137, and the Starting Research Fund from
Guangzhou University under Grant No.
RQ 2020083.

\section*{Author Declarations}
The authors have no conflicts to disclose.

\appendix
%
%

\section{Proof of the integral identity Eq.(\ref{Identity})}
We have
\begin{eqnarray}
   & & \int_0^\infty e^{-R^2t^2- \alpha^2/4t^2 } dt  \nonumber \\
   &=& \int_0^\infty e^{-x^2 - \alpha^2 R^2/4x^2 } \frac{dx }{R} \nonumber \\
   &=& \int_0^\infty e^{-(x - \alpha R/2 x )^2 - \alpha R } \frac{dx }{R} \nonumber \\
   &=& \frac{e^{-\alpha R}}{R} \int_0^\infty e^{-(x- \alpha R/2x)^2} dx .
\end{eqnarray}
Next we prove that for arbitrary $A >0 $,
\begin{eqnarray}
  \frac{\sqrt{\pi}}{2} =I=  \int_0^\infty e^{-(x- A/x)^2 } d x .
\end{eqnarray}
With the change of variable $y = A/x$, we have
\begin{eqnarray}
  &I  = \int_0^\infty e^{-(y-A/y)^2} \frac{A dy }{y^2} \notag\\
  &=\int_0^\infty e^{-(x-A/x)^2} d\left(-\frac{A}{x} \right ).
\end{eqnarray}
Therefore,
\begin{eqnarray}
  &I =\frac{1}{2}  \int_0^\infty e^{-(x-A/x)^2} d\left( x -\frac{A}{x} \right ) \notag\\
  &=  \frac{1}{2}\int_{-\infty}^\infty e^{-z^2} dz =  \frac{1}{2}\sqrt{\pi} .
\end{eqnarray}

\section{Calculation  of Eq.(\ref{error})}
We have the integral
\begin{eqnarray}\label{int1}
 I = \frac{2}{\sqrt{\pi }} \int_\eta^\infty \exp\left(-A^2 t^2 + \frac{B^2}{4t^2}\right ) dt ,
\end{eqnarray}
where $A>0 $ and $B>0 $. Since the above integral makes sense for arbitrary complex number $B$, in the end, we shall analytically continue the result to an arbitrary complex value of $B$.
Introducing the variable $z_\pm = At \pm iB/2t$, we have
\begin{eqnarray}\label{int}
  &&I =\notag\\
  && \frac{2}{\sqrt{\pi } }\frac{1}{2A} \{ \int_{A\eta + i B/2 \eta }^\infty e^{-z_+^2 + i AB } d z_+\!\! + \!\!\int_{A\eta -i B/2 \eta }^\infty e^{-z_-^2 - i AB } d z_- \} \notag\\
  &&= \frac{1}{2A} \left[ e^{i AB } erfc(A\eta + i B/2 \eta)  + e^{-iAB} erfc(A\eta - i B/2 \eta)  \right ].\notag\\
\end{eqnarray}
Here we have the error function and the complementary error function
\begin{eqnarray}
erf(x) &=&  \frac{2}{\sqrt{\pi } } \int_0^x e^{-z^2 } dz, \nonumber \\
  erfc(x) &=& 1- erf(x) =   \frac{2}{\sqrt{\pi } } \int_x^\infty e^{-z^2 } dz.
\end{eqnarray}
Note that $erf(x)$ is an entire and odd function of $x$.
The final result (\ref{int}) is analytic in $B$, and then it makes sense also for purely imaginary $B$, e.g., for negative energy $B^2=E=q^2<0$ in Eq.(\ref{161}).

 \section{the vanishing of wave function at origin for any non-trivial irreducible representation of  point group}
In this appendix, we prove that if an irreducible unitary representation is not a trivial representation of point group $G(\mathbf{k})$  (the group of wave vector $\mathbf{k}$), its basis function  would be zero at origin.
We assume that the basis function of a $j-$th irreducible representation is $\psi^{j}_m(\mathbf{r})$, the irreducible  (unitary) representation matrix is $(D^{j})_{l_j\times l_j}$ where $l_j$ is dimension of representation space.
For a group operation $R$ acting on spatial point $\textbf{r}$, the change of  basis function is given by
\begin{align}\label{A1}
P_R \psi^{j}_m(\mathbf{r})=\psi^{j}_m(R^{-1}\mathbf{r})=\psi^{j}_n(\mathbf{r})D^{j}_{n,m}(R),
\end{align}
where $P_R$ is corresponding operator of group operation $R$.

For a point group $G(\mathbf{k})$, because the origin is a fixed, i.e., $R^{-1}\mathbf{r}=R^{-1}\mathbf{0}=\mathbf{0}$, Eq.(\ref{A1}) is reduced into
\begin{align}\label{A2}
&\psi^{j}_m(\mathbf{0})=\psi^{j}_n(\mathbf{0})D^{j}_{n,m}.
\end{align}
Summing over all the group operation $R$ in Eq.(\ref{A2}), we get
\begin{align}\label{A3}
&g\psi^{j}_m(\mathbf{0})=\psi^{j}_n(\mathbf{0})M_{n,m}\notag\\
&M_{n,m}\equiv\sum_{R}D^{j}_{n,m}(R)=\sum_{R}(1)_{1, 1}\times D^{j}_{n,m}(R),
\end{align}
where integer $g$ is the order of group $G(\mathbf{k})$. $1=(1)_{1,1}$  can be viewed as matrix element of the one dimension trivial representation matrix $(1)_{1\times 1}$ of the group of wave vector.
In the Eq.(\ref{A3}), we introduce matrix $M$.
Because  $D^{j}(R)$ is a non-trivial irreducible representation matrix, based on the orthogonality condition of irreducible representation matrices \cite{Tung},
all the matrix elements of $M$ should be zero, i.e., $M_{n,m}=0$.
From Eq.(\ref{A3}), we get that all  the basis functions at origin take zero values, i.e., $\psi^{j}_m(\mathbf{0})=0$.

If the basis functions are divergent near origin , e.g., the wave function  $\psi^{j}_m(\mathbf{r})\propto A_m/r$ as $r\rightarrow 0$ [see Eq.(\ref{201}], where $A_m$ is a constant,  Eq.(\ref{A1}) can  be multiplied by a factor $r$, and then one can use a similar procedure to  get that $A_m=0$ and $\psi^{j}_m(r)=0$ as $r\rightarrow0$.

We note that, for other continuous compact Lie groups, e.g., SU(2) or SO(3) group, if one uses an integration over continuous group elements to replace summation over finite group elements in Eq.(\ref{A3}),  a similar conclusion that the basis functions at origin of non-trivial irreducible representation are zero, can be obtained with an exactly same method.

\section{Proof of $[E+d^2/dx^2]e^{iq|x-n|}=2iq\delta(x-n)$}
The above equation can be obtained as following.
Using $E=q^2$, we known that if $x>n$
\begin{align}
&[E+d^2/dx^2]e^{iq|x-n|}=[E+d^2/dx^2]e^{iq(x-n)}\notag\\
&=[q^2+(iq)^2]e^{iq(x-n)}=0,
\end{align}
and if $x<n$ 
\begin{align}
&[E+d^2/dx^2]e^{iq|x-n|}=[E+d^2/dx^2]e^{-iq(x-n)}\notag\\
&=[q^2+(-iq)^2]e^{-iq(x-n)}=0.
\end{align}
So when $x\neq n$,
 \begin{align}
[E+d^2/dx^2]e^{iq|x-n|}=0.
\end{align}
However due to the singularity (or non-analyticity) of absolute value function $|x-n|$ near $x=n$,  $[E+d^2/dx^2]e^{iq|x-n|}$ may be proportional to Dirac's delta function.
So we can assume that 
\begin{align}
[E+d^2/dx^2]e^{iq|x-n|}=A \delta(x-n),
\end{align}
where $A$ is a constant.
In a small interval near $x=n$, we integrate above equation, i.e., 
\begin{align}
&\int^{n+0^+}_{n-0^+} dx [E+d^2/dx^2]e^{iq|x-n|}\notag\\
&=[e^{iq|x-n|}]'|_{x=n+0^+}-[e^{iq|x-n|}]'|_{x=n-0^+}\notag\\
&=[e^{iq(x-n)}]'|_{x=n+0^+}-[e^{-iq(x-n)}]'|_{x=n-0^+}\notag\\
&=[iqe^{iq(x-n)}]|_{x=n+0^+}-[-iqe^{-iq(x-n)}]|_{x=n-0^+}\notag\\
&=2iq=\int^{n+0^+}_{n-0^+} dx A \delta(x-n)=A,
\end{align}
then we get $A=2iq$ and 
\begin{align}
[E+d^2/dx^2]e^{iq|x-n|}=2iq \delta(x-n).
\end{align}
We also notice that $e^{iq|x-n|}/(2iq)$ is a Green's function of one dimensional free particle with Hamiltonian $H_0=-d^2/dx^2$ (see Ref. \cite{Economou2006}).

\end{document}